# TOPOLOGICAL FIELD THEORY AS THE KEY TO QUANTUM GRAVITY

by: Louis Crane, mathematics department,KSU

**Abstract**: Motivated by the similarity between CSW theory and the Chern Simons state for General Relativity in the Ashtekar variables, we explore what a universe would look like if it were in a state corresponding to a 3D TQFT. We end up with a construction of propagating states for parts of the universe and a Hilbert space corresponding to a certain approximation. The construction avoids path integrals, using instead recombination diagrams in a very special tensor category.

## 1. INTRODUCTION

What I wish to propose is that the quantum theory of gravity can be constructed from a topological field theory. Notice, I do not say that it **is** a topological field theory. Physicists will be quick to point out that there are local excitations in gravity, so it cannot be topological.

Nevertheless, the connections between general relativity in the loop formulation and the CSW TQFT are very suggestive. Formally, at least, the CSW action term

*

gives a state for the Ashtekar variable version of quantized general relativity, as we have heard at this conference.

That is, we can think of this "measure" on the space of connections on a 3 manifold either as defining a theory in 3 dimensions, or as a state in a 4-D theory, which is GR with a cosmological constant.

Furthermore, states for the Ashtekar or loop variable form of GR are very scarce, unless one is prepared to consider states of zero volume. Finally, one should note that a number of very talented workers in the field have searched in vain for a Hilbert space of states in either representation.

If one attempts to use the loop variables to quantize subsystems of the universe, one ends up looking at functionals on the set of links in a 3 manifold with boundary with ends on the boundary. It turns out in regularizing that the links need to be framed, and one can trace them in any representation of SU(2), so they need to be labeled. The coincidence of all this with the picture of 3D TQFT was the initial motivation for this work.

Another point, which seems important, is that although path integrals in general do not exist, the CSW path integral can be thought of as a symbolic shorthand for a 3-D TQFT which can be rigorously defined by combining combinatorial topology with some very new algebraic structures, called modular tensor categories. Thus, if we want to explore * as a state for GR, we have the possibility of switching from the language of path integrals to a finite, discrete, picture, where physics is described by variables on the edges of a triangulation.

Physics on a lattice is nothing new. What is new here is that the choice of lattice is unimportant, because the coefficients attached to simplices are algebraically special. Thus we do not have to worry about a continuum limit, since a finite result is exact.

Motivated by these considerations, I propose to consider the hypothesis:

CONJECTURE: The universe is in the CSW state.

This idea is similar to the suggestion of Witten that the CSW theory is the topological phase of quantum gravity [2]. My suggestion differs from his in assuming that the universe is still in the topological phase, and that the concrete geometry we see is not the result of a fluctuation of the state of the universe as a whole, but rather due to the collapse of the wave packet in the presence of classical observers. As I shall explain below, the machinery of CSW theory provides spaces of states for parts of the universe, which I interpret as the relative states an observer might see. This suggestion is closely related to the many worlds interpretation of quantum mechanics: the states on surfaces are data in one particular world. The state of the universe is the sum of all possible worlds, which takes the elegant form of the CSW invariant of links, or more concretely, of the Jones polynomial or one of its generalisations.

Durring the talk I gave at the conference one questioner pointed out, quite correctly, that it is not necessary to make this conjecture, since other states exist. Let me emphasize that it is only a hypothesis. I think it is fair to say that it leads to more beautiful mathematics than any other I know of. Whether that is a good guide for physics only time will tell.

Topological quantum field theory in dimensions 2,3, and 4 is leading into an extraordinarily rich mathematical field. In the spirit of the drunkard who looks for his keys near a street light, I have been thinking for several years about an attempt to unite the structures of TQFT with a suitable reinterpretation of quantum mechanics for the entire universe. I believe that the two subjects resemble one another strongly enough that the program I am outlining becomes natural.

## 2. QUANTUM MECHANICS OF THE UNIVERSE. CATEGORICAL PHYSICS.

I want to propose that a quantum field theory is the wrong structure to describe the entire universe. Quantum field theory, like quantum mechanics generally, presupposes an external classical observer. There can be no probability interpretation for the whole universe. In fact, the state of the universe cannot change, because there is no time, except in the presence of an external clock. Since the universe as a whole is in a fixed state, the quantum theory of the gravitational field as a whole is not a physical theory. It describes all possible universes, while the task of a physical theory is to explain measurements made in **this** universe in **this** state. Thus, what is needed is a theory which describes a universe in a particular state. This is similar to what some researchers, such as Penrose, have suggested: that the initial conditions of the universe are determined by the laws. What I am proposing is that the initial conditions become part of the laws. Naturally, this requires that the universe be in a very special state.

Philosophically, this is similar to the idea of a wave function of the universe. One can phrase this proposal as the suggestion that the wave function of the universe is the CSW functional. There is also a good deal here philosophically in common with the recent paper of Rovelli and Smolin [23]. Although they do not assume the universe is in the CSW state, they couple the gravitational field to a particular matter field, which acts as a clock, so that the gravitational field is treated as only part of a universe.

What replaces a quantum field theory is a family of quantum theories corresponding to parts of the universe. When we divide the universe into two parts, we obtain a Hilbert space which is associated to the boundary between them. This is another departure from quantum field theory; it amounts to abandonment of observation at a distance.

The quantum theories of different parts of the universe are not, of course, independent. In a situation where one observer watches another there must be maps from the space of states which one observer sees to the other. Furthermore, these maps must be consistent. If A watches B watch C watch the rest of the universe, A must see B see C see what A sees C see.

Since, in the Ashtekar/loop variables the states of quantum gravity are invariants of embedded graphs, we allow labeled punctures on the boundaries between parts of the universe, and include embedded graphs in the parts of the universe we study, ie. in 3 manifolds with boundary.

The structure we arrive at has a natural categorical flavor. Objects are places where observations take place, ie boundaries; and morphisms are 3D cobordisms, which we think of as A observing B.

Let us formalize this as follows:

Definition: **An <u>Observer</u> is an oriented 3 manifold with boundary containing an embedded labeled framed graph which intersects the boundary in isolated labeled points.**

(The labeling sets for the edges and vertices of the graph are finite, and need to be chosen for all of what ensues.)

Definition: **A <u>skin of observation</u> is a closed oriented surface with labeled punctures.**

Definition: **If A and B are skins of observation an <u>inspection, $\alpha$, of B by A</u> is an observer whose boundary is identified with $\overline{A} \cup B$ (i.e. reverse the orientation on A) such that the labelings of the components of the graph which reach the boundary of $\alpha$ match the labelings in $\overline{A} \cup B$.**

This definition requires that the set of labelings possess an involution corresponding to reversal of orientation on a surface.

To every inspection $\alpha$ of B by A there corresponds a dual inspection of $\overline{A}$ by $\overline{B}$, given by reversing orientation and dualizing on $\alpha$.

Definition: **The <u>category of observation</u> is the category whose objects are skins of observation and whose morphisms are inspections.**

Definition: **If $M^3$ is a closed oriented 3-manifold the <u>category of observation in $M^3$</u> is the relative (i.e. embedded) version of the above.**

Of course, we can speak of observers, etc. in $M^3$.

Definition: **A state for quantum gravity (in $M^3$) is a functor from the category of observation (in $M^3$) to the category of vector spaces.**

Nowhere in any of this do we assume that these boundaries are connected. In fact, the most important situation to study may be one in which the universe is crammed full of a "gas" of classical observers. We shall discuss this situation below as a key to a physical interpretation of our theory.

If we examine the mathematical structure necessary to produce a "state" for the universe in this sense, we find that it is identical to a 3D TQFT. Thus, the CSW state produces a state in this new sense as well.

Another way to look at this proposal is as follows: The CSW state for the Ashtekar variables is a very special state, in that it factorizes when we cut the 3 manifold along a surface with punctures, so that a finite dimensional Hilbert space is attached to each such surface, and the invariant of any link cut by the surface can be expressed as the inner product of two vectors corresponding to the two half links. (John Baez has pointed out that these finite dimensional spaces really do possess natural inner products). Thus, it produces a "state" also in the sense we have defined above.

I interpret this as saying that the state of the universe is unchanging, but that because the universe is in a very special state, it can contain a classical world, ie. a family of classical observers with consistent mutual observations. The states on pieces of the universe (ie. links with ends in manifolds with boundary) can be interpreted as changing, once we learn to interpret vectors in the Hilbert spaces as clocks.

So far, we have a net of finite dimensional Hilbert spaces, rather than one big one, and no idea how to reintroduce time in the presence of observers. There are natural things to try for both of these problems in the mathematical context of TQFT.

Before going into a program for solving these problems (and thereby opening the possibility of computing the results of real experiments) let us make a survey of the mathematical toolkit we inherit from TQFT.

3. IDEAS FROM TQFT

As we have indicated, the notion of a "state of quantum gravity" we defined above is equivalent to the notion of a 3D TQFT which is currently prevalent in the literature.

The simplest definition of a 3D TQFT is that it is a machine which assigns a vector space to a surface, and a linear map to a cobordism between 2 surfaces. The empty surface receives a 1 dimensional vector space, so a closed 3 manifold gets a numerical invariant. Composition of cobordisms corresponds to compo-sition of the linear maps. A more abstract way to phrase this is that a TQFT is a functor from the cobordism category to the category of vector spaces.

The TQFTs which have appeared lately are richer than this. They assign vector spaces to surfaces with labeled punctures, and linear maps to cobordisms containing links or knotted graphs with labelled edges. The labels correspond to representations. Since it is easy to extend the loop variables to allow traces in arbitrary representations (characters), there is a great deal of coincidence in the pictures of CSW TQFT and the loop variables. (That was a lot of the initial motivation for this program). We can describe this as a functor from a richer cobordism category to VECT. The objects in the richer cobordism category are surfaces with labelled punctures, and the morphisms are cobordisms containing labelled links with ends on the punctures.

There are several ways to construct TQFTs in various dimensions. Let us here discuss the construction of a TQFT from a triangulation. We assign labels to edges in the triangulation, and some sort of factors combining the edge labels to different dimensional simplices in the triangulation, multiply the factors together, and sum over labelings.

In order to obtain a topologically invariant theory, we need the combination factors to satisfy some equations. The equations they need to satisfy are very algebraic in nature; as we go through different classes of theories in different low dimensions we first rediscover most of the interesting classes of associative algebras, then of tensor categories [19].

A simple example in two dimensions may explain why the equations for a topological theory have a fundamental algebraic flavor. For a 2D TQFT defined from a triangulation, we need invariance under the move:

. Now, if we think of the coefficients which we use to combine the labels around a triangle as the structure coefficients of an algebra, this is exactly the associative law. Much of the rest of classical abstract algebra makes an appearance here too.

As has been described elsewhere [3], a 3D TQFT can be constructed from a modular tensor category. The interesting examples of MTCs can be realized as quotients of the categories of representations of quantum groups.

If we try to use the modular tensor categories to construct a TQFT on a triangulation, we obtain, not the CSW theory, but the weaker TQFT of Viro and Turaev [20]. Here we label edges, not from a basis for an algebra, but with irreducible objects from a tensor category. The combinations we attach to tetrahedra (3 simplices) are the quantum 6J symbols, which come from the associativity isomorphisms of the category.

We refer to this sort of formula as a state summation.

The full CSW theory can also be reproduced from a modular tensor category by a slightly subtler construction which uses a Heegaard splitting (which can be easily produced from a triangulation) [3].

The quantum 6J symbols satisfy some identities, which imply that the summation formula for a 3 manifold is independent under a change of the triangulation. The most interesting of these is the Eliot-Biedenhorn identity. This identity is the consistency relation for the associativity isomorphism of the MTC. this is another example of the marriage between algebra and topology which underlies TQFT.

The suggestion that this sort of summation could be related to the quantum theory of gravity is older than one might think. If we use the representations of an ordinary lie group instead of a quantum group, then the Viro-Turaev formula becomes the Regge-Ponzano formula for the evaluation of a spin network.

Regge and Ponzano [9], were able to interpret the evaluation as a sort of discrete path integral for $3 - d$ euclidean quantum gravity. The formula which Regge and Ponzano found for the evaluation of a graph is the analog for a lie group of the state sum of Viro and Turaev for a quantum group. Thus, the evaluation of a tetrahedron for a spin network is a 6J symbol for the group SU(2).

(1)

The topological invariance of this formula follows from some elementary properties of representation theory. In (1), we have placed our trivalent labeled graph on the boundary of a 3-manifold with boundary, then cut the interior up into tetrahedra, labeling the edges of all the internal lines with arbitrary spins. (The Clebsch-Gordon Relations imply that only finitely many terms in (1) are non-zero.)

Regge and Ponzano then proceeded to interpret (1) as a discretized path integral for 3D quantum GR. The geometric interpretation consisted inthinking of the Casimirs of the representations as lengths of edges. Representation theory implied that the summation was dominated by flat geometries [9].

Another way to think of the program in this paper is an attempt to find the appropriate algebraic structure for extending the spin network approach to quantum gravity from D=3 to D=4.

Another idea from TQFT, which seems to have relevance for this physical program is the ladder of dimensions [19]. TQFTs in adjacent dimensions seem to be related algebraically. The spirit of the relationship is like the relationship of a tensor category to an algebra. Tensor categories look just like algebras with the operation symbols in circles. The identities of an algebra correspond to isomorphisms in a tensor category. These isomorphisms then satisfy higher order consistency or"coherence" relations, which relate to topology up one dimension.

The program of construction in TQFT, which I hope will yield the tools for the quantization of general relativity, is not yet completed, but is progressing rapidly. There are two outstanding problems, which are mathematically natural, and which I believe are crucial to the physical problem as well.They are:

Topological problem 1: find a triangulation version of CSW theory (not merely its absolute square as in [20])

Topological problem 2: Construct a 4D TQFT related via the dimensional ladder to CSW theory.

The solution of these two problems will be in reach, if the program in [19] succeeds. The program of [19] also suggests that the solutions of these two topological problems are closely related, both being constructed from state sums involving the same new algebraic tools.

In this context we should also note that a 4D TQFT has been constructed in [21], out of a modular tensor category. This construction begins with a triangulation of a 4 manifold, and picks a Heegaard splitting for the boundary of each 4-simplex of the triangulation. Thus the theory in [21] can be thought of as producing a 4D theory by filling the 4 space with a network of 3D subspaces containing observers. This connection between TQFTs in 3D and 4D is the sort of thing I believe we will need to solve the problem of reintroducing time in a universe in a TQFT state. What I expect is that the program of [19] will provide richer examples of such a connection, which will prove to be relevant to quantizing General Relativity.

What we conjecture in [19], is that a new algebraic structure will give us a new state summation, similar to the one we discussed above, which will give us CSW in three dimensions. If we use the new summation in four dimensions, we should obtain a discrete version of $F \wedge F$ theory. This combination seems very suggestive, since we are supposed to be obtaining our relative states from CSW theory on a boundary, while $F \wedge F$ theory is a lagrangian for the topological sector for the quantum theory of GR [22]. Also. the algebraic structure we need to construct seems to be an expanded quantum version of the lorentz group. It is this new, not yet understood summation, which comes from the representation theory of the new structure which we call an F algebra in [19], which I believe should give us a discretized version of a path integral for quantum gravity.

4.HILBERT SPACE IS DEAD. LONG LIVE HILBERT SPACE. or THE OBSERVER GAS APPROXIMATION.

Let us assume that we have found a suitable state summation formula to solve our 2 topological problems (so that what we suggest will be predicated on the success of the program in [19], although one could also try to use the state sum in [21] together with Viro-Turaev theory). There is a natural proposal to make a physical interpretation within it in a 4 dimensional setting. In the scheme I am suggesting, we can recover a Hilbert space in a certain sort of classical limit, in which the universe is full of a "gas," of classical observers. The idea is that if we choose a triangulation for the 4 manifold and a Heegaard splitting for the boundary of each 4 simplex, then we obtain a family of 2-surfaces which can be thought of as filling up the 4 manifold.These surfaces should be thought of as skins of observation for a family of observers who are crowding the spacetime. Now let us think of a 4 manifold with boundary. In a three dimensional boundary component, we can then combine the vector spaces which we assign to the surfaces which cut the boundaries of those 4 simplices lying on the boundary into a larger Hilbert space, which is a quotient of their tensor product. (It is a quotient because the surfaces overlap.) Now, if we pass to a finer triangulation, we will obtain a larger space, with a linear map to the smaller one. The linear map comes from the fact that we are using a 3D TQFT, and the existence of 3D cobordisms connecting pieces of the surfaces of the two sets of Heegaard splittings.

We end up with a large vector space for each triangulation of the boundary 3 manifold, and a linear map when one triangulation is a refinement of another. This produces a directed graph of vector spaces and linear maps associated to a 3 manifold. The 4D state sum can be extended to act on the vectors in these vector spaces by extending a triangulation for the boundary 3 manifold to one for the 4 manifold.

Whenever we have a directed set of vector spaces and consistent linear maps, there is a construction called the inverse limit which can combine them into a single vector space. The vectors are vectors in any of the spaces, with images under the maps identified. Since the 4D state sum is assumed to be topologically invariant, we can extend the linear map it assigns to a 4D cobordism to a map on the inverse limit spaces. I propose that it is these inverse limit spaces which play the role of the physical Hilbert spaces of the theory.

The thought here is that physical Hilbert space is the union of the finite dimensional spaces which a gas of observers can see, in the limit of an infinitely dense gas.

The effect of this suggestion is to reverse the relationship between the finite dimensional spaces of states which each observer can actually see and the global hilbert space. We are regarding the states which can be observed at one skin of observation as primary. This "relational" approach bears some resemblance to the physical ideas of Leibnitz, although that is not an argument for it.

5. THE PROBLEM OF TIME

The test of this proposal is whether it can reproduce Einstein's equations in some classical limit. What I am proposing is that the 4D state sum which I hope to construct from the representation theory of the F algebra can be interpreted as a discretized version of the path integral for general relativity. The key question is whether the state sum is dominated in the limit of large spins on edges by assignments of spins whose geometric interpretation would give a discrete approximation to an Einstein manifold. This would mean that the theory was a quantum theory whose classical limit was general relativity.

The analogy with the work of Regge and Ponzano is the first suggestion that this program might work. In [9], they interpreted the summation we wrote down above as a discretized path integral for 3D quantum gravity. The geometric interpretation consisted in thinking of the Casimirs of representations as lengths. It was somewhat easier to get Einstein's equations in D=3, since the solutions are just flat metrics.

Should this program work? One objection, which was raised in the discussion, is that the $F \wedge F$ model may only be a sector of quantum gravity in some weak sense. (I am restating this objection somewhat in the absence of the questioner). It can be replied that since the theory we are writing has the right symmetries, the action of the renormalization group should fix the lagrangian. I do not think that either the objection or the reply is decisive in the abstract. If the state sum suggested in [19] can be defined, then we can investigate whether we recover Einstein's equation or not.

If this does work, one could do the state sum on a 4 manifold with corners, ie, fix states on surfaces in the three dimensional boundaries of a 4D cobordism. One could interpret these states as initial conditions for an experiment, and investigate the results by studying the propagation via the 4D state sum.

One could conjecture that this would give a spacetime picture for the evolution of relative states within the framework of the CSW state of the universe.

## 6. MATTER AND SYMMETRY

It is natural to ask whether this picture could be expanded to include matter fields. The obvious thing to try is to pass from SU(2) to a larger group. It may be noted that Peldan [12] has observed that the CSW state also satisfies the Yang Mills equation. If we really get a picture of gravity from the state sum associated to SU(2), it would not be a great leap to try a larger gauge group.

As I was writing this I became aware of [24], in which Gambini and Pullin point out that the CSW state is also a state for GR coupled to electromagnetism. They also say that the sources for such a theory, thought of as places where links have ends, are necessarily fermionic. These results seem to further strengthen the program set forth here.

In general, the constructions which lead up to the F algebras in [19] can be thought of as expressions of "quantum symmetry" ie as relatives of lie groups in a quantum context. For example, the modular tensor categories can be described as deformed Clebsch Gordon coefficients. The F algebras arise as a result of pushing this process further, recategorifying the categories into 2-categories. These constructions can also be done from extremely simple starting points, looking at representations of Dynkin diagrams as "quivers." The idea that theories in physics should be reconstructed from symmetries actually originated with Einstein, who was thinking of general relativity. It would be fitting somehow, to reconstruct truly fundamental theories of physics from fundamental mathematical expressions of symmetries.

**Acknowledgements:** the author wishes to thank Lee Smolin and Carlo Rovelli for many years of conversation on this extremely elusive topic. The mathematical ideas here are largely the result of collaborations with David Yetter and Igor Frenkel. Louis Kauffman has provided many crucial insights. The author is supported by NSF grant DMS-9106476 .References.

1. C. Rovelli and L. Smolin, Loop Space Representations of Quantum General Relativity, preprint IV 20 Phys. Dept. V. di Roma "la Sapienza" 1988

2. E. Witten, Quantum Field Theory and The Jones Polynomial, IAS preprint HEP 88/33

3. L. Crane, 2-d Physics and 3-d Topology, Commun. Math. Phys. 135 615-640 (1991)

4. G. Moore and N. Seiberg, Classical and Quantum Conformal Field Theory, Commun Math. Phys. 123 177-254 (1989)

5. B. Brugman, R Gambini and J. Pullin, Gen. Rel. Grav. in press

6. L. Crane, Quantum Symmetry, Link Invariants and Quantum Geometry, in Proceedings XX International Conference on differential Geometric Methods in Theoretical Physics Baruch College (1991)

7. L. Crane, Conformal Field Theory, Spin Geometry, and Quantum Gravity, Phys. Lett. B v259 #3 (1991)

8. R. Penrose, Angular Momentum; an Approach to Combinatorial Space Time, in Quantum Theory and Beyond ed T.Bastin(Cambridge)

9. G. Ponzano and T. Regge, Semiclassical Limits of Racah Coefficients, in Spectroscopic and Group Theoretical Methods in Physics, ed. F. Bloch (North-Holland, Amsterdam)

10. C. Rovelli, Class. Quan. Grav. 8 297 (1991)

11. J. Moussouris, Quantum Models of Space Time Based on Recoupling Theory, Thesis, Oxford University (1983)

12. P. Peldan, Phys. Rev. D46 R2279 (1992)

13. L. Smolin, Personal communication

14. V. Moncrief, personal communication

15. K. Walker, On Wittens 3 Manifold Invariants, unpublished

16. L. Crane and Igor Frenkel, Hopf Categories and Their Representations, to appear.

17. M. Kapranov and V. Voevodsky, Braided Monoidal 2-Categories, 2-Vector Spaces and Zamolodchikov's Tetrahedra Equation, preprint

18. C. Rovelli and L. Smolin, Phys. Rev. Lett. 61 1155 (1988)

19. L. Crane and I. Frenkel, A Representation theoretic Approach to 4-Dimensional Topological Quantum Field Theory, to appear in proceedings of AMS conference, Mt. Holyoke MA., June 1992

20 V. Turaev and O. Viro, State Sum Invariants of 3-Manifolds and Quantum 6J Symbols Topology 31 (1992) 865-902

21 L. Crane and D. Yetter, A Categorical Construction of 4D TQFT's To appear in Proceedings of AMS conference, Dayton, Ohio, October, 1992

22 L. N. Chang and C. Soo, BRST Cohomology and Invariants of 4D Gravity in Ashtekar Variables VPI preprint10

# FOUR DIMENSIONAL TQFT; a Triptych

by Louis Crane[1]
Mathematics Department
Kansas State University
Manhattan KS, 66506

ABSTRACT: We discuss three possibly interrelated ideas. One idea is a suggestion for a categorical approach to quantizing gravity, one tells us how to lift a TQFT from D=3 to D=4, and one tells us how to refine some modular tensor categories into 2-categories. Relations between the ideas are conjectured.

Since it is impossible to state one idea adequately in 20 minutes, I shall outline three ideas sketchily instead. I am submitting to this conference, as joint work with David Yetter, a preliminary version of a paper which details one of the three ideas [1]. A second is written out in [2], and a third in [3].

The three ideas I describe begin in different places, but ought to connect with one another. They all revolve around the problem of constructing Topological Quantum Field Theory [TQFT] in dimension 4.

*1 Quantum Gravity and augmented TQFT*

There seems to be a connection between 3D TQFTs and the quantum theory of gravity, as formulated in the Ashtekar variables. There are several indications of this. One is that the Chern-Simons functional, which Witten [4] used to formally construct 3D TQFT, is (also formally), a solution of the constraints for general relativity in the Ashtekar variables [5]. Another indication is the deformed spin network interpretation of the CSW TQFT, [6] which shows that it is closely related to the spin network formulation of 3D quantum gravity due to Regge and Ponzano [7].

The deformed spin network picture gives us a picture of a 3D TQFT as a rule which assigns to a 3-manifold with boundary a number, which can be interpreted as an average over discretized 3D Riemannian metrics, when we specify a state in the space of conformal blocks of the boundary surface. This goes a long way towards giving an interpretation of the CSW TQFT as a state for the quantum theory of gravity.

In order to create an interpretation for the CSW state which would be related to physical experiments, it is necessary to reintroduce time into our picture, i.e. to make a 4-dimensional structure which is some sort of extension of a 3d TQFT. The word I coin for the new structure is "augmented."

An analysis of what would be physically useful for a 4D interpretation suggests a mathematically elegant form for what an augmentation should be. A

[1] supported by NSF grant DMS-9106476

TQFT can be described as a functor from a cobordism category to Vect (the category of finite-dimensional vector spaces) It is easy to restrict a TQFT to subsets of some 3-manifold M and get a functor on a relative cobordism category, of surfaces and cobordisms embedded in M.

An augmentation of a TQFT then amounts to a rule, which assigns, to any 4D cobordism between two 3-manifolds a natural transformation between the two functors associated to the relative categories of the two 3-manifolds. This translates physically into time evolution operators on relative states, together with appropriate consistency relations when observers observe one another evolving.

This leads us to the problem of trying to augment a 3D TQFT. Since CSW theory can be constructed from a modular tensor category (MTC) [8], we are led to consider two separate mathematical problems :

1) augmenting a 3D TQFT directly up to D=4, and

2) augmenting a modular tensor category into a 2-category.

I do not, at this point know how to use either of these ideas to construct an augmentation in the sense described above, but there is considerable progress to report on both, purely as mathematical problems.

*2. Categorical construction of 4D TQFT.*

This construction is discussed in the short paper submitted to the conference jointly with D. Yetter.Consequently, I confine myself here to a few simple remarks.

One important point is that we have a construction of a 4D TQFT from a type of tensor category we are already familiar with, rather than a hypothetical one which solves some list of axioms.

It is rather surprising that we get a 4D theory from a modular tensor category, since the general intuition in the field is that 4D invariants should come from 2-categories [9, 10]. The construction has another rather surprising feature, which is that we put labels of spins (or, more generally, of representations) on the 2-simplices of the triangulation rather than on the edges.

We believe that these two facts suggest that our invariant is a special case of one constructed from 2-categories. The MTC is playing the role of a 2-category with one object, so that the edges are labeled trivially, and the representations on the surfaces are really to be thought of as morphisms.

As I shall mention below, this suggests that the 2-category I describe in section 3 may have some relationship to 4D topology.

*3. The Canonical Basis and 2-Categories*

This part of my talk represents joint work with Igor Frenkel [3].

There are several ways to construct the MTCs which are related to CSW theory. One is to look at the representation theory of quantum groups when q is a root of unity. Quantum groups admit a rigid and elegant layer of structure related to the canonical basis of Lusztig [11], in which the structure constants and tensor operators between representations have purely integral coefficients, and with slight modification of the algebra, even purely positive integral coefficients.

At first glance, the structure related to the canonical basis seems unrelated to the topological applications of quantum groups. We do not know if we can use the canonical basis to produce 4D invariants, but we do have a result which strongly suggests that possibility.

Our result is that the quantum groups can be "categorified", ie. represented as the Grothendieck rings of tensor categories. The categories of representations of the quantum groups can then be recategorified into 2-categories.

We believe that this "categorification" is indicative of possible 4D topological significance because of an analogous phenomenon which relates 2D and 3D TQFTs via categorification.

(A detailed discussion of the notion of categorification can be found in [3]. Let me just explain, briefly, that if we have a category with direct sum and tensor product, we can make a ring of the formal linear combinations of objects in the obvious way. This we call the Grothendieck ring. The reverse operation, i.e. from ring to category, is called "categorification". It can be nonexistent or many-valued. Clearly, a ring needs a basis in which its operations have positive integral structure constants in order to be categorified. That is the point of departure for the connection between the canonical basis and categorification.)

2D TQFTs are constructible from commutative associative algebras with an inner product and suitable compatibility conditions. On the other hand, 3D TQFTs are constructed from modular tensor categories.

It turns out that the most physically interesting 2D TQFTs, namely the G/G models, are constructed from the algebras constructed from the chiral vertex operators of WZW models, called Verlinde algebras. The Verlinde algebras are precisely the Grothendieck algebras of the modular tensor categories which are related to CSW theories. Thus the 3D CSW theories are categorifications of the G/G models.

As yet, it is only a matter of conjecture that the 2-category we construct has any relationship with the 4D construction in part 2.

4. *Connections*

It is not yet clear if the three ideas outlined above connect. Recently, L. Smolin [12] has suggested that in order to make a 4-dimensional interpretation of states for general relativity in the Ashtekar variables, it is necessary to fill up space with a family of surfaces on which we could locate clocks.

Since the 4D construction described in 2 above is really built up out of 3D TQFT on a family of 2-surfaces in the 3-skeleton, it may be possible to interpret our 4D construction as implementing Smolin's suggestion under the assumption that the universe is in the CSW state. If so, and if a steepest descent approximation exists for our 4D theory analogous to Regge-Ponzano theory in D=3, then the relationship between ideas 1 and 2 in this paper may be close.

Acknowledgements:

Much of this work is joint work with D. Yetter and I Frenkel. the ideas about quantum gravity grow out of years of work with L. Smolin and C. Rovelli. I wish to thank J. Baez, J. Pullin, and G. Zuckerman for helpful conversations.